# Quantum geometry quadrupole-induced third-order nonlinear transport in antiferromagnetic topological insulator MnBi$_2$Te$_4$

Hui Li[1#], Chengping Zhang[1#], Chengjie Zhou[1], Chen Ma[1], Xiao Lei[2], Zijing Jin[1], Hongtao He[2], Baikui Li[1], Kam Tuen Law[1]*, Jiannong Wang[1]*

[1]*Department of Physics, the Hong Kong University of Science and Technology, Clear Water Bay, Hong Kong, China*

[2]*Department of Physics, South University of Science and Technology of China, Shenzhen, Guangdong 518055, China*

[#]These authors contributed equally to this work.

*Correspondence and requests for materials should be addressed to K. T. L. (email: phlaw@ust.hk), and J. W. (email: phjwang@ust.hk).

**Abstract.** The study of quantum geometry effects in materials has been one of the most important research directions in recent decades. The quantum geometry of a material is characterized by the quantum geometric tensor of the Bloch states. The imaginary part of the quantum geometry tensor gives rise to the Berry curvature while the real part gives rise to the quantum metric. While Berry curvature has been well studied in the past decades, the experimental investigation on the quantum metric effects is only at its infancy stage. In this work, we measure the nonlinear transport of bulk MnBi$_2$Te$_4$, which is a topological anti-ferromagnet. We found that the second order nonlinear responses are negligible as required by inversion symmetry, the third-order nonlinear responses are finite. The measured third-harmonic longitudinal ($V_{xx}^{3\omega}$) and transverse ($V_{xy}^{3\omega}$) voltages with frequency $3\omega$, driven by an a.c. current with frequency $\omega$, show an intimate connection with magnetic transitions of MnBi$_2$Te$_4$ flakes. Their magnitudes change abruptly as MnBi$_2$Te$_4$ flakes go through magnetic transitions from an antiferromagnetic state to a canted antiferromagnetic state and to a ferromagnetic state. In addition, the measured $V_{xx}^{3\omega}$ is an even function of the applied magnetic field $\boldsymbol{B}$ while $V_{xy}^{3\omega}$ is odd in $\boldsymbol{B}$. Amazingly, the field dependence of the third-order responses as a function of the magnetic field suggests that $V_{xx}^{3\omega}$ is induced by the quantum metric quadrupole and $V_{xy}^{3\omega}$ is induced by the Berry curvature quadrupole.

Therefore, the quadrupoles of both the real and the imaginary part of the quantum geometry tensor of bulk $MnBi_2Te_4$ are revealed through the third order nonlinear transport measurements. This work greatly advanced our understanding on the connections between the higher order moments of quantum geometry and nonlinear transport.

The study of the intricate relationship between quantum geometric properties and high-order nonlinear transport properties in topological materials is at the frontier of condensed matter physics research [1-16]. The quantum geometric properties of a material are characterized by the quantum geometric tensor $T_{\alpha\beta}(\boldsymbol{k}) = \langle\partial_\alpha u_k|(1 - |u_k\rangle\langle u_k|)|\partial_\beta u_k\rangle$. Here, $|u_k\rangle$ denotes a Bloch state with crystal momentum $\boldsymbol{k}$ and $\partial_\alpha$ denotes the derivative with respect to the $\alpha$-component of $\boldsymbol{k}$. The imaginary part of $T$, $\Omega_{\alpha\beta} = 2\text{Im}(T_{\alpha\beta})$, defines the Berry curvature and the real part $g_{\alpha\beta} = \text{Re}(T_{\alpha\beta})$ defines the quantum metric. Importantly, it has been shown that the Berry curvature, the quantum metric and their dipole distribution can induce novel nonlinear transport phenomena [8,13,15-25]. For instance, the Berry curvature dipole induced second-order nonlinear Hall effect has been well demonstrated in bilayer and few-layer $WTe_2$ flakes [5,17,26-29]. More recently, with $MnBi_2Te_4$ thin films, quantum metric dipole induced second-order nonlinear Hall effect was observed in which an alternating current (a.c.) with frequency $\omega$ induces an a.c. Hall voltage with frequency $2\omega$ [20,21]. However, inversion symmetry must be broken in these $MnBi_2Te_4$ films such that the second-order effect can be observed.

In this work, we study the higher-order nonlinear transport properties of bulk $MnBi_2Te_4$ which is a topological anti-ferromagnet. As expected, due to the presence of the inversion symmetry in the bulk, both the second order Hall response and the second order longitudinal response are zero, which is different from the observations in thin films [20,21]. Incredibly, we observed a third order longitudinal response in which an a.c. current with frequency $\omega$ induces an a.c. voltage $V_{xx}^{3\omega}$ with frequency $3\omega$ along the current direction. Interestingly, in the presence of an out-of-plane magnetic field, an a.c. current induces an a.c. voltage $V_{xy}^{3\omega}$ with frequency $3\omega$ in the transverse direction. Through the symmetry analysis and the semi-classical Boltzmann equation approach, we show that $V_{xx}^{3\omega}$ is induced by the quantum metric quadrupole $\partial_x^2 G_{xx}$ (where $G_{\alpha\beta}$ is the normalized quantum metric defined in the Method Section). On the other hand, the third order Hall voltage $V_{xy}^{3\omega}$ is induced by the Berry curvature quadrupole which was predicted recently [13]. More importantly, the observed magnetic field dependence of $V_{xx}^{3\omega}$ and $V_{xy}^{3\omega}$ match the calculated third order response well.

In other words, in bulk MnBi$_2$Te$_4$, the suppression of the second order nonlinear transport by inversion symmetry enabled the third order nonlinear response to be the leading nonlinear response, which is induced by the quadrupole distribution of the Berry curvature and the quantum metric. This is the first experimental work revealing the higher moments of both the real and the imaginary part of the quantum metric tensor and deepened our understanding of the relation between quantum geometry, crystal symmetry and the nonlinear transport.

Here, we explain the origin of the quantum geometry quadrupole of a material and its nonlinear transport properties. In particular, the distribution of Berry curvature in a genetic time-reversal broken system could result in the formation of non-zero Berry curvature quadrupoles even the Berry curvature monopole and the dipole are zero as depicted in Fig. 1(a). When an electric field is introduced, we reach a steady state in which the electric field induces a Berry curvature dipole along the electric field direction, as demonstrated in Fig. 1(b). This electric field-induced dipole will cause a nonlinear response in the third order of the electric field as explained in Ref.13. This Berry curvature quadrupole induced third order effect offers complementary experimental knobs to investigate the high-order nonlinear anomalous Hall effect in magnetic materials. Analogously, a finite normalized quantum metric $G_{\alpha\beta}$ can induce a finite Berry connection and $\partial_x^2 G_{xx}$ gives rise to the third order nonlinear response [15]. As we show below, bulk antiferromagnetic topological insulator MnBi$_2$Te$_4$ is an ideal platform for investigating the quantum geometry quadrupole effects as the second harmonics are suppressed by inversion symmetry and the third order harmonics are the leading order nonlinear response.

Antiferromagnetic topological insulator MnBi$_2$Te$_4$ has rich magnetic structures and striking topological band structures [7, 30-40]. The lattice structure of MnBi$_2$Te$_4$ is shown in Fig. 1(c). It has a septuple layer structure in each unit cell, which composes a Bi$_2$Te$_3$ quintuple layer and an MnTe layer. Each Mn layer in the unit cell contributes a typical ferromagnetic feature, while the adjacent two Mn layers are antiferromagnetic coupled at low temperatures. The bulk MnBi$_2$Te$_4$ preserves an inversion symmetry ($\mathcal{P}$) and a time-reversal times half-translation symmetry ($\mathcal{T}t_{1/2}$), giving rise to a combined $\mathcal{PT}$ symmetry. The $\mathcal{PT}$ symmetry will force the Berry curvature quadrupoles to vanish. However, when a magnetic field (**B**-field) is applied to induce a finite magnetization, the $\mathcal{PT}$ symmetry will be broken and a nonzero Berry curvature quadrupole will emerge, which can induce the third order nonlinear Hall effect [13, 22]. On the other hand, the quantum metric-induced nonlinear effects are allowed by the $\mathcal{PT}$ symmetry [11], therefore can be observed even in the absence of the external **B**-fields. In recent experiments [20, 21] on thin films, inversion symmetry is broken and the second-order nonlinear effects were observed. In contrast, the bulk $MnBi_2Te_4$ preserves the inversion symmetry, making the third-order nonlinear effects the leading-order contributions from the quantum metrics.

In this work, we experimentally investigate the third-order nonlinear transport properties in bulk $MnBi_2Te_4$ flakes. At temperatures below the antiferromagnetic transition temperature of $MnBi_2Te_4$ flakes, both the third-harmonic longitudinal ($V_{xx}^{3\omega}$) and transverse ($V_{xy}^{3\omega}$) voltages are observed. These high-order nonlinear responses are tightly related to magnetic transitions of $MnBi_2Te_4$ flakes. In addition, the measured $V_{xx}^{3\omega}$ and $V_{xy}^{3\omega}$ show even and odd dependence on the external **B**-fields respectively, due to the distinct physical origins from the quantum metric and the Berry curvature quadrupoles. Importantly, the observed **B**-field dependence of the $V_{xx}^{3\omega}$ and $V_{xy}^{3\omega}$ can be well-explained by the calculations from an effective four-band Hamiltonian which respects the symmetry properties of the crystal.

**Results**

**Temperature dependence of nonlinear transport properties of $MnBi_2Te_4$ flakes.** Fig. 2(a) shows the schematic drawing of the measurement setup for the nonlinear transport properties of $MnBi_2Te_4$ flakes. The a.c. with a low frequency of 179 Hz was injected and the longitudinal voltage ($V_{xx}$) and transverse voltage ($V_{xy}$) signals at different harmonics were collected using lock-in techniques. The thickness of $MnBi_2Te_4$ flakes is ~60 nm as determined by the atomic force microscope measurement in Fig. S1 in the Supplementary Information. Striking third-harmonic longitudinal and transverse voltage are clearly observed in $MnBi_2Te_4$ flakes with their magnitude being increased with increasing first-harmonic current, as shown in Fig. S2 in the Supplementary Information. On the contrary, no distinct second-harmonic signal has been observed in $MnBi_2Te_4$ flakes due to the crystallographic symmetry constraint (Fig. S3 in the Supplementary Information). We note that the magnitude of the third-harmonic signal of a resistor of ~100 $\Omega$ is negligible. This rules out the instrumental errors as a possible origin of the observed third-harmonic signal in MnBi$_2$Te$_4$ flakes.

Fig. 2(b) shows the temperature ($T$) dependence of the first-harmonic longitudinal resistance ($R_{xx}^{\omega}$) of the MnBi$_2$Te$_4$ flakes collected at different out-of-plane $\boldsymbol{B}$-fields ($\boldsymbol{B} \perp \boldsymbol{I}$) with an a.c. current of 0.6 mA. As temperature decreases, the resistance decreases gradually with an abnormal peak occurring at $T_N$ ~21.7 K, corresponding to the antiferromagnetic (AFM) transition in MnBi$_2$Te$_4$ flakes [30,37]. It is noted that the measured AFM transition temperature $T_N$ ~21.7 K is slightly lower than the measured $T_N$ of ~23.7 K with the applied current of ~1 μA (Supplementary Fig. S4(a)) and previous studies of ~24 K [30,37]. After safely excluding the heating effect [41], such deviation is probably due to the large current caused suppression of the antiparallel alignments of the spins in MnBi$_2$Te$_4$ flakes and the spin-transfer torque effect [42,43]. In addition, the AFM transition in MnBi$_2$Te$_4$ flakes is strongly related to out-of-plane $\boldsymbol{B}$-fields. As indicated by the orange curve in Fig. 2(b), the AFM transition in MnBi$_2$Te$_4$ flakes shifts to low temperatures of ~19 K at 2 T. Further increasing the $\boldsymbol{B}$-fields to 4 T, a striking resistance upturn occurs at temperature below ~18 K and gradually saturates at temperature below ~5 K, as indicated by the blue curve in Fig. 2(b). Such an intriguing transition is believed to be related to the formation of a canted AFM state in MnBi$_2$Te$_4$ flakes at low temperatures [37]. At higher $\boldsymbol{B}$-fields above 7 T, the AFM transition in MnBi$_2$Te$_4$ flakes smears out with resistance continuously decreases with decreasing temperature in the entire temperature ranging from 100 K to 2 K, behaving as a metal (green curve for 7 T and purple curve for 9.5 T) [36,37].

The temperature dependence of third-harmonic longitudinal ($V_{xx}^{3\omega}$) and transverse ($V_{xy}^{3\omega}$) voltage signals of MnBi$_2$Te$_4$ flakes collected at different out-of-plane $\boldsymbol{B}$-fields are shown in the upper and lower panel of Fig. 2(c), respectively. The $V_{xx}^{3\omega}$ ($V_{xy}^{3\omega}$) of MnBi$_2$Te$_4$ flakes is obtained by averaging of the sum (difference) of voltages measured at positive and negative $\boldsymbol{B}$-fields, respectively. Both $V_{xx}^{3\omega}$ and $V_{xy}^{3\omega}$ are strongly dependent on the $\boldsymbol{B}$-fields. As it can be seen from the upper panel of Fig. 2(c), at 0 T, the $V_{xx}^{3\omega}$ decreases slowly with decreasing temperature from 70 K to ~37 K. Below ~37 K, it starts to increase and reach a peak value at ~23.7 K, then decreases sharply (black curve). The rapid change around 23.7 K coincides with the AFM transition of MnBi$_2$Te$_4$ flakes, suggesting the change of $V_{xx}^{3\omega}$ may relate to the AFM transition of MnBi$_2$Te$_4$ flakes. Further decreasing temperature, the $V_{xx}^{3\omega}$ of MnBi$_2$Te$_4$ flakes features a slight upturn at ~18.7 K followed by a continuous decrease at temperature below ~13.7 K. These striking features remain unchanged as **B**-field increases to 2 T except the transition temperature shifted to lower temperature by ~2 K comparing with that at 0 T (orange curve). While for higher **B**-field of 4 T, the $V_{xx}^{3\omega}$ gradually decreases with decreasing temperature from 70 K to 35 K, then increases as temperature below 35 K, and finally saturates with temperature below ~5 K (blue curve). In contrast, the $V_{xx}^{3\omega}$ continuously decreases with decreasing temperature from 70 K to 2 K for the large **B**-field of 7 T (green curve) and 9.5 T (purple curve).

The temperature dependence of $V_{xy}^{3\omega}$ of MnBi$_2$Te$_4$ flakes measured at different **B**-fields is shown in the lower panel of Fig. 2(c). At **B**-fields of 0 T (black curve) and 2 T (orange curve), the $V_{xy}^{3\omega}$ is almost zero at temperatures far above the AFM transition temperature of MnBi$_2$Te$_4$ flakes with a slight increase with decreasing temperature (see zoom-in view in the inset of lower panel). As temperature approaches the AFM transition temperature of MnBi$_2$Te$_4$ flakes, an abrupt change of $V_{xy}^{3\omega}$ occurs within a narrow temperature range from ~23 K to ~19.5 K at **B** = 0 T (black curve), where its magnitude changes by ~18 μV from ~3 μV to ~ -15 μV. Further decreasing temperature, the $V_{xy}^{3\omega}$ shows a slight upturn followed by a gradual decrease. Similar abrupt changes near the AFM transition of MnBi$_2$Te$_4$ flakes are also observed at **B** = 2 T (orange curve) and 4 T (blue curve). However, the magnitude of the $V_{xy}^{3\omega}$ is greatly enhanced at high **B**-fields of 4 T. In contrast, the $V_{xy}^{3\omega}$ exhibits a continuous increase with decreasing temperatures at even higher **B**-fields of 7 T (green curve) and 9.5 T (purple curve). Notably, the magnitude of $V_{xy}^{3\omega}$ reaches ~65 μV at 2 K and 9.5 T, which is almost one order of magnitude larger than that at 0 T.

**B-fields dependence of nonlinear transport properties in MnBi$_2$Te$_4$ flakes.** Figure 3(a) shows the out-of-plane **B**-fields dependent $V_{xx}^{3\omega}$ of MnBi$_2$Te$_4$ flakes measured at different temperatures. The $V_{xx}^{3\omega}$ curves in Fig. 3(a) are symmetrized by averaging the sum of magnitude collected in the positive and negative **B**-fields regions to exclude any mixture effect. It is clear that all curves exhibit a distinct even-symmetry with **B**-fields directions. Such an even-symmetry of the $V_{xx}^{3\omega}$ is well evidenced even from the raw data in Fig. S5(a) in the Supplementary Information. The evolution of $V_{xx}^{3\omega}$ is strongly associated with the magnetic transitions of MnBi$_2$Te$_4$ flakes under applied **B**-fields. The MnBi$_2$Te$_4$ flakes undergo multiple magnetic transitions from AFM state to canted-AFM state and then to polarized ferromagnetic (FM) state with increasing **B**-fields, as illustrated by the magnetoresistance curve of MnBi$_2$Te$_4$ flakes at 2 K in Supplementary Fig. S4(b) in the Supplementary Information. At 2 K and small **B**-fields (< ~2.7 T) with MnBi$_2$Te$_4$ flakes in the AFM state, the $V_{xx}^{3\omega}$ is almost unchanged (black curve). With increasing **B**-fields, a sudden drop changing from ~-230 μV to ~ -460 μV followed by an abrupt flipping from ~ -460 μV to ~175 μV occurs in a narrow **B**-field range from ~2.7 T to ~3.2 T. This corresponds to the magnetic transition from AFM state to canted AFM state in MnBi$_2$Te$_4$ flakes, as indicated in Supplementary Fig. S4(b) in the Supplementary Information. Further increasing **B**-fields, the $V_{xx}^{3\omega}$ decreases continuously through the magnetic transition from canted-AFM state to FM state. Finally, it saturates as MnBi$_2$Te$_4$ flakes enter the polarized FM state at higher **B**-field above ~8 T. Similar behaviors are observed at temperatures below AFM transition temperature of MnBi$_2$Te$_4$ flakes, as indicated by the red curve for 10 K, blue curve for 15 K, and green curve for 20 K, respectively, in Fig. 3(a). However, as temperature increases above the AFM transition temperature of MnBi$_2$Te$_4$ flakes, these abrupt changes in the $V_{xx}^{3\omega}$ disappear. As a result, the $V_{xx}^{3\omega}$ only shows a slight and monotonical change with increasing **B**-fields, for example, the $V_{xx}^{3\omega}$ slightly changes by ~30 μV as **B**-fields increase from 0 to 9 T at 30 K (purple curve) and is almost unchanged at 50 K (brown curve). Further increasing the temperature to 100 K, the magnitude of $V_{xx}^{3\omega}$ only shows a slight increase with increasing **B**-fields (cyan curve).

The out-of-plane **B**-fields dependent $V_{xy}^{3\omega}$ of MnBi$_2$Te$_4$ flakes at different temperatures are shown in Fig. 4(a). The $V_{xy}^{3\omega}$ curves are symmetrized by averaging the magnitude difference between the positive and negative **B**-fields regions to exclude the mixing of $V_{xx}^{3\omega}$. The $V_{xy}^{3\omega}$ is of odd-symmetry, which is in striking contrast to that of $V_{xx}^{3\omega}$, as demonstrated in Fig. 4(a) and the raw data in Fig. S5(b) in the Supplementary Information. The evolution of $V_{xy}^{3\omega}$ is also strongly related to the magnetic transitions of the MnBi$_2$Te$_4$ flakes, like that in the longitudinal direction. At a low temperature of 2 K (black curve), the magnitude of $V_{xy}^{3\omega}$ closes to 0 and is almost unchanged in the AFM state of MnBi$_2$Te$_4$ flakes in small $\boldsymbol{B}$-fields < ~2.7 T. With increasing $\boldsymbol{B}$-fields, the magnitude of $V_{xy}^{3\omega}$ decreases sharply by ~50 μV in a narrow $\boldsymbol{B}$-fields ranging from ~2.7 T to ~3.2 T. This coincides with the abrupt change occurring in the $V_{xx}^{3\omega}$ (Fig. 3(a)), which is related to the magnetic transition from AFM state to canted AFM state of MnBi$_2$Te$_4$ flakes in the intermediate $\boldsymbol{B}$-fields [36,37]. Further increasing $\boldsymbol{B}$-fields from ~3.2 T to ~7.5 T, the $V_{xy}^{3\omega}$ increases continuously through the magnetic transition from canted AFM state to FM state in MnBi$_2$Te$_4$ flakes. Then, the $V_{xy}^{3\omega}$ saturates at even higher $\boldsymbol{B}$-fields above 7.5 T, corresponding to the formation of the polarized FM state in MnBi$_2$Te$_4$ flakes. The overall features of $V_{xy}^{3\omega}$ are the same at temperatures below AFM transition temperatures except that the fields corresponding to the sharp decrease shift to lower $\boldsymbol{B}$-fields with increased temperatures, as demonstrated in Fig. 4(a) for 10 K (red curve), 15 K (blue curve) and 20 K (green curve), respectively. In contrast, the $V_{xy}^{3\omega}$ features a monotonic increase as temperature further increases above the AFM transition temperature of MnBi$_2$Te$_4$ flakes. The magnitude of $V_{xy}^{3\omega}$ increases by ~22.5 μV with increasing $\boldsymbol{B}$-fields from 0 to 9 T at 30 K (purple curve), and becomes temperature independent with further increasing temperatures, as shown by the brown and cyan curves at 50 K and 100 K, respectively.

**Discussion** The distinct symmetry properties of the Berry curvature and quantum metric contributions allow us to distinguish the two effects as both are observed in our experiments. The third-order nonlinear response contributed by the Berry curvature quadrupoles is odd under the time-reversal symmetry. Furthermore, the Berry curvature quadrupoles only contribute to the third-harmonic Hall voltage $V_{xy}^{3\omega}$ without longitudinal contributions, so it can account for the observed $V_{xy}^{3\omega}$ in MnBi$_2$Te$_4$ flakes with odd-symmetry with respect to the $\boldsymbol{B}$-fields directions. On the other hand, the third-order nonlinear response contributed by the quantum metric is even under the time-reversal symmetry and only contributes to the longitudinal third-harmonic voltage $V_{xx}^{3\omega}$, therefore, it is responsible for the observed $V_{xx}^{3\omega}$ in MnBi$_2$Te$_4$ flakes with even-symmetry with respect to the $\boldsymbol{B}$-fields directions.

In order to capture the effect induced by the out-of-pane $\boldsymbol{B}$-field, we introduce a model Hamiltonian with a tunable magnetic order parameter $\Delta$ (see Methods for details). Figure 3(b) plots the calculated quantum metric induced third-harmonic conductivity $\sigma_{xxxx}$ as a function of the magnetic order parameter $\Delta$, see calculation details in Methods. As it can be seen, the $\sigma_{xxxx}$ is strongly dependent on $\Delta$ and its variation with $\Delta$ reproduces the key features of our experimental measured $\boldsymbol{B}$-fields dependence of $V_{xx}^{3\omega}$. Figures 3(c-e) illustrate the physical origin of the evolution of the $\sigma_{xxxx}$ with increasing $\Delta$. At zero $\boldsymbol{B}$-field or $\Delta = 0$, all bands of MnBi$_2$Te$_4$ are doubly degenerate due to the combined $\mathcal{PT}$ symmetry (Fig. 3(c)). Each band contributes a finite quantum metric with its sign opposites for conduction band (positive) and valance band (negative), as shown in Fig. 3(c). Such finite quantum metric will result in the observation of finite third-harmonic voltage in the longitudinal direction $V_{xx}^{3\omega}$, as observed in the experimental measurements in Fig. 3(a). When applying a $\boldsymbol{B}$-field or $\Delta \neq 0$, the broken $\mathcal{PT}$ symmetry in MnBi$_2$Te$_4$ will split the degenerated bands and modify the distribution of the quantum metric, as illustrated in Fig. 3(d). The changes of the quantum metric distribution will lead to the observation of abrupt changes in $V_{xx}^{3\omega}$, as plotted in Fig. 3(b) in the intermediate magnetic order parameters and evidenced from the experimental measurements in the transition from the AFM state to canted-AFM state in MnBi$_2$Te$_4$ in Fig. 3(a). Further increasing the magnetic order parameters, the splitting between bands becomes larger, as shown in Fig. 3(e). This will modulate the quantum geometric properties of the electrons close to Fermi surfaces and lead to the occurrences of sharp change, as demonstrated by theoretical calculation in Fig. 3(b) and experimental observations in Fig. 3(a).

On the other hand, the experimentally observed odd-symmetry of $V_{xy}^{3\omega}$ can be explained by the nonzero Berry curvature quadrupoles when the $\mathcal{PT}$ symmetry is broken by the non-zero magnetic order parameter $\Delta$ or the external $\boldsymbol{B}$-fields. Fig. 4(b) shows the calculated magnetic order parameter $\Delta$ dependence of Berry curvature quadrupoles in MnBi$_2$Te$_4$. Remarkably, it exhibits a good agreement with our experimental observations in Fig. 4(a). Figs. 4(c-e) show the calculated band structure evolution and the strength distribution of the Berry curvature quadrupole. The combined $\mathcal{PT}$ symmetry of MnBi$_2$Te$_4$ flakes forces each band to be doubly degenerated in the absence of external $\boldsymbol{B}$-fields, as indicated by the black curve in Fig. 4(c). This band degeneracy makes the Berry curvature quadrupoles to be zero, resulting in the zero $V_{xy}^{3\omega}$ in the antiferromagnetic state of MnBi$_2$Te$_4$ flakes. When the external $\boldsymbol{B}$-field is applied, the $\mathcal{PT}$ symmetry is broken lifting the band degeneracy. Thus, a non-zero Berry curvature quadrupole emerges, as illustrated in Fig. 4(d). This coincides with our experimental observations in Fig. 4(a), where an abrupt change occurs in the transition from AFM state to canted-AFM state at increased $\boldsymbol{B}$-fields. With further increasing the magnetic order parameter $\Delta$, the band splitting enhances and eventually pushes the upper band away from the Fermi level (Fig. 4(e)). Consequently, the sharp change of the $V_{xy}^{3\omega}$ occurs again, as shown in the theoretical calculation in Fig. 4(b) and the experimental observations in Fig. 4(a) at larger $\boldsymbol{B}$-fields. The detailed theoretical derivations are presented in Methods.

## Conclusion

In conclusion, our systematic study of nonlinear transport properties of the antiferromagnetic topological insulator MnBi$_2$Te$_4$ flakes has evidenced the existence of third-harmonic voltages in both longitudinal and transverse directions at temperatures below the antiferromagnetic transition temperature of MnBi$_2$Te$_4$ flakes. The observed third-order nonlinear transport properties are related to the magnetism of MnBi$_2$Te$_4$ flakes with the distinct changes of the third-harmonic voltages closely following the magnetic transitions. The longitudinal third-harmonic response $V_{xx}^{3\omega}$ is attributed to the quantum metric origin. In contrast, the transverse third-harmonic response $V_{xy}^{3\omega}$ is originated from the emergency of non-zero Berry curvature quadrupoles induced by the external $\boldsymbol{B}$-fields. Our findings provide the connection between the quantum geometry quadrupole and the high-order nonlinear transport properties while lower order nonlinear responses are suppressed by inversion symmetry. We expect that even higher order (such as the seventh order) responses, can be the leading order nonlinear responses given the appropriate symmetry [13]. This work opens a new venue to explore quantum geometry multipoles induced nonlinear Hall effect.

## Methods

**Devices fabrication and magnetoresistance measurements.** To study the magnetotransport properties, a Hall-bar geometry was patterned through standard electron-beam lithography and lift-off techniques. Au/Cr electrodes with thickness of 100 nm/10 nm were deposited using thermal evaporation methods. The magnetotransport properties of the devices were then measured in a Quantum Design Physical Property Measurement System using standard lock-in technique.

**Theoretical calculations of third-order nonlinear conductivity.**

The generic third-order conductivity can be defined as $j_\alpha = \sigma_{\alpha\beta\gamma\delta} E_\beta E_\gamma E_\delta$. In particular, the third-order longitudinal conductivity induced by the quantum metric is [15]

$$\sigma_{xxxx} = \frac{e^4\tau}{\hbar^2}\left(-\int_k \partial_x^2 G_{xx} f_0 + \frac{1}{2}\int_k v_x^2 G_{xx} f_0''\right),$$

Where $\boldsymbol{v} = \boldsymbol{\nabla}_k \varepsilon_k$ is the band velocity, $\partial_x \equiv \partial/\partial_{k_x}$, and $G_{ab} = 2Re\sum_{m\neq n}\frac{(\mathcal{A}_a)_{nm}(\mathcal{A}_b)_{mn}}{\varepsilon_n - \varepsilon_m}$ is the band-energy normalized quantum metric tensor [2, 11, 20, 21]. In order to get a gauge-independent result, we have used the relationship $(\mathcal{A}_a)_{nm}(\mathcal{A}_b)_{mn} = \frac{\langle n|\partial_a \mathcal{H}|m\rangle\langle m|\partial_b \mathcal{H}|n\rangle}{(\varepsilon_n - \varepsilon_m)^2}$ in our calculations.

On the other hand, the third-order Hall conductivity induced by the Berry curvature quadrupole is [13]

$$\sigma_{yxxx} = \frac{e^4\tau^2}{\hbar^3} Q_{xxz},$$

where $Q_{xxz} = \int_k \partial_x^2 \Omega_z f_0$ is the Berry curvature quadrupole.

Next, we introduce the effective Hamiltonian of MnBi$_2$Te$_4$ flakes in our theoretical calculations for third-order nonlinear effects induced by the quantum metric and Berry curvature quadrupole. Ref. [44] provides a model Hamiltonian of ferromagnetic MnBi$_2$Te$_4$, where the ferromagnetic state is stabilized by the external $\boldsymbol{B}$-field. Here, in order to study the effect with the increase of the $\boldsymbol{B}$-field, we introduce a tunable magnetic order parameter $\Delta$, and the effective Hamiltonian reads,

$$\mathcal{H}(\boldsymbol{k}) = \mathcal{H}_0(\boldsymbol{k}) + \Delta \, \mathcal{H}_{\mathrm{M}}(\boldsymbol{k}),$$

where $\mathcal{H}_0(\boldsymbol{k})$ respects the $\mathcal{PT}$ symmetry, and $\mathcal{H}_{\mathrm{M}}(\boldsymbol{k})$ is the part which is induced by the finite magnetization and breaks the time-reversal symmetry. The detailed parameters of the Hamiltonian can be found in Supplementary Note 5.

## Acknowledgements

This work was supported in part by the Research Grants Council of the Hong Kong SAR under Grant Nos. 16306421, AoE/P-701/20-3, and C6025-19G.

## Author Contributions

J.N.W. conceived the research; C.M. and X.L. fabricated devices with support from H.H.; H.L. performed transport experiments with support from C.Z., C.M., Z.J., B.L.; C.Z and K.T.L. provided theoretical support; H.L., C.P., H.H., K.T.L. and J.N.W. analyzed experimental data and wrote the manuscript with input from all co-authors.

## Additional information

**Competing financial interests:** The authors declare no competing financial interests.

## Figure Captions

**Figure 1. Berry curvature quadrupole in MiBi₂Te₄.** (a) Schematic diagrams of the band structure (the outline) and Berry curvature (color scale) of a generic magnetic material with non-zero Berry curvature quadrupoles. (b) Electric field induced change of Berry curvature relative to the equilibrium state. Due to the Berry curvature distribution of the quadrupole, a field-induced dipole will appear which contributes to the higher order nonlinear anomalous Hall effect. (c) Schematic crystal structure of layered MiBi₂Te₄. The red arrows in Mn layers indicate the spin direction of Mn atoms.

**Figure 2. Temperature dependence of the nonlinear transport properties in MiBi₂Te₄.** (a) Schematic drawing of the measurement setup. (b) Temperature dependent of the first-harmonic longitudinal resistance at different out-of-plane $\boldsymbol{B}$-fields with excited a.c. current of 0.6 mA. (c) Temperature dependence of the $V_{xx}^{3\omega}$ (upper panel) and $V_{xy}^{3\omega}$ (lower panel) at different out-of-plane $\boldsymbol{B}$-fields with excited a.c. current of 0.6 mA. Inset in lower panel shows the zoom-in view of $V_{xy}^{3\omega}$ at low temperatures.

**Figure 3. $\boldsymbol{B}$-fields dependence of the longitudinal third-harmonic voltage in MiBi₂Te₄.** (a) The out-of-plane $\boldsymbol{B}$-fields dependence of the $V_{xx}^{3\omega}$ measured at different temperatures. (b) The calculated $\sigma_{xxxx}$ contributed by quantum metric as a function of the magnetic order parameter $\Delta$. (c-e) The band splitting and distribution of quantum metric $G_{xx}$ under different magnetic order parameter $\Delta = 0$ (c), $\Delta = 0.2$ (d), and $\Delta = 0.7$ (e), where the black dashed line indicates the chemical potential.

**Figure 4. $\boldsymbol{B}$-fields dependence of the transverse third-harmonic voltage in MiBi₂Te₄.** (a) The out-of-plane $\boldsymbol{B}$-fields dependence of the $V_{xy}^{3\omega}$ measured at different temperatures. (b) The calculated Berry curvature quadrupole $Q_{xxz}$ as a function of the magnetic order parameter $\Delta$. (c-e) The band splitting and distribution of Berry curvature quadrupole $\Omega_z$ at different magnetic order parameter $\Delta = 0$ (c), $\Delta = 0.2$ (d), and $\Delta = 0.7$ (e).

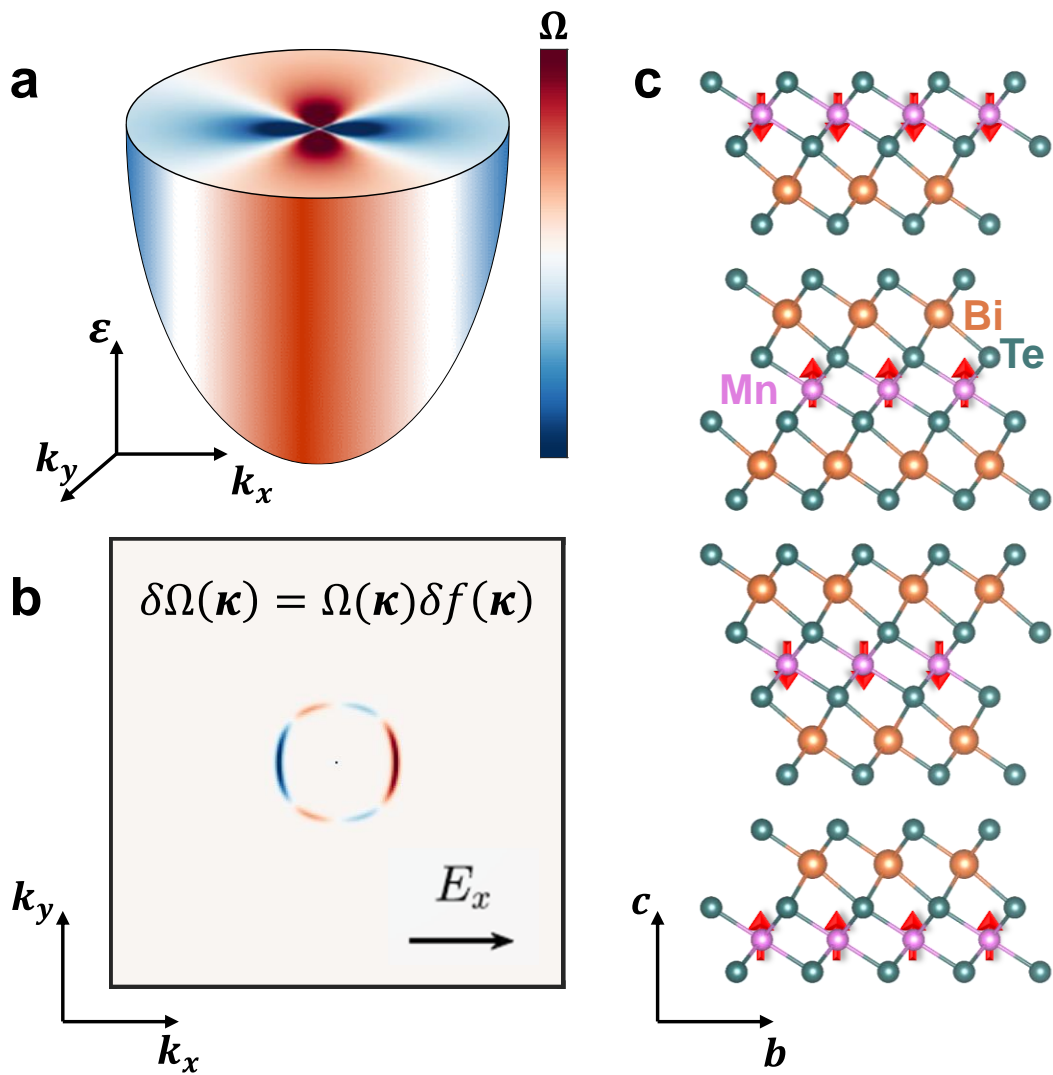

**a**

$\Omega$

$\varepsilon$

$k_y$

$k_x$

**b**

$\delta\Omega(\boldsymbol{\kappa}) = \Omega(\boldsymbol{\kappa})\delta f(\boldsymbol{\kappa})$

$k_y$

$k_x$

$E_x$

**c**

Bi

Te

Mn

$c$

$b$

Figure 1

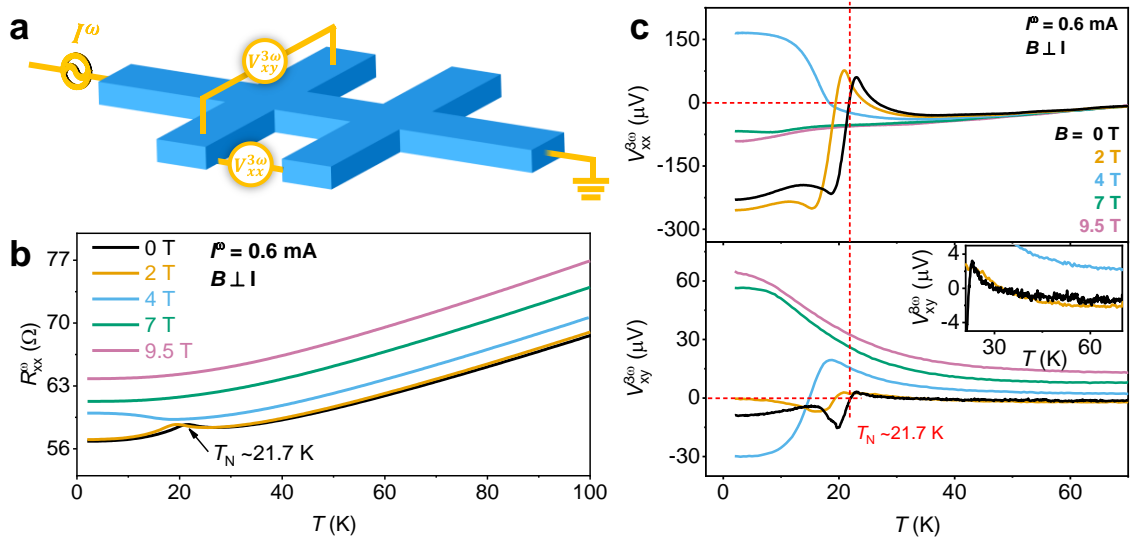

**Figure 2**

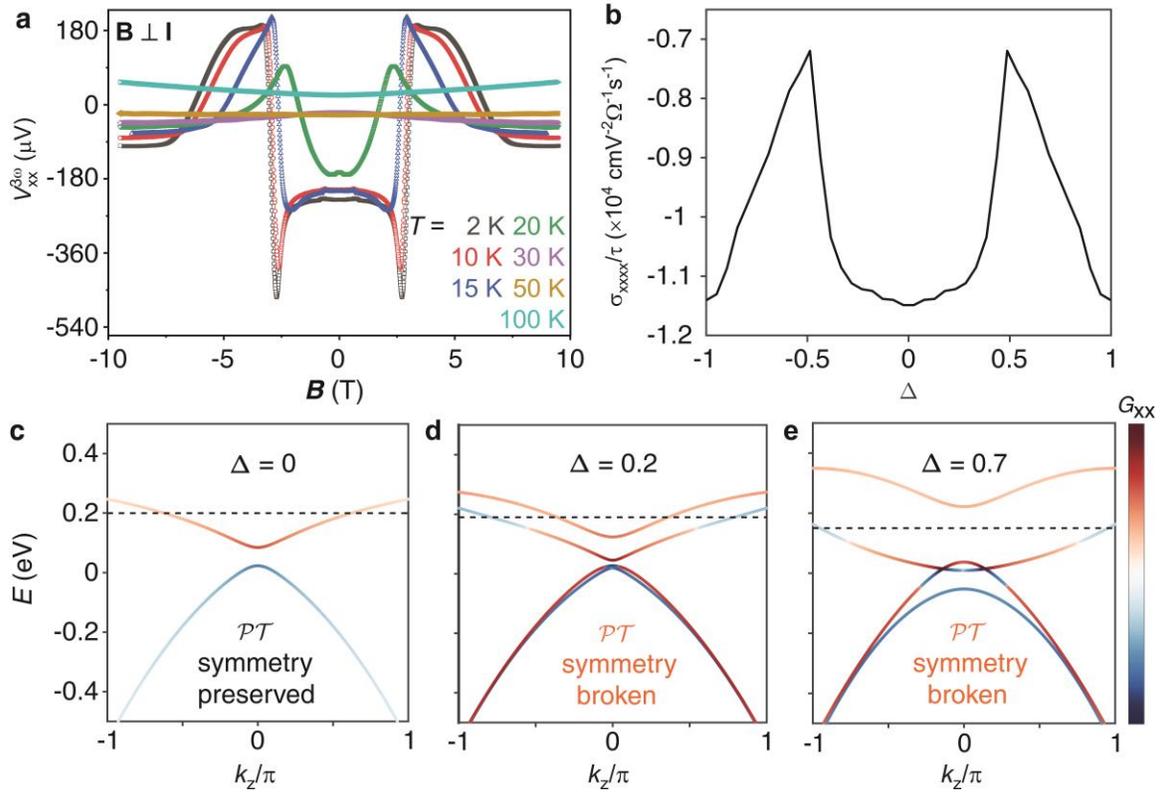

**Figure 3**

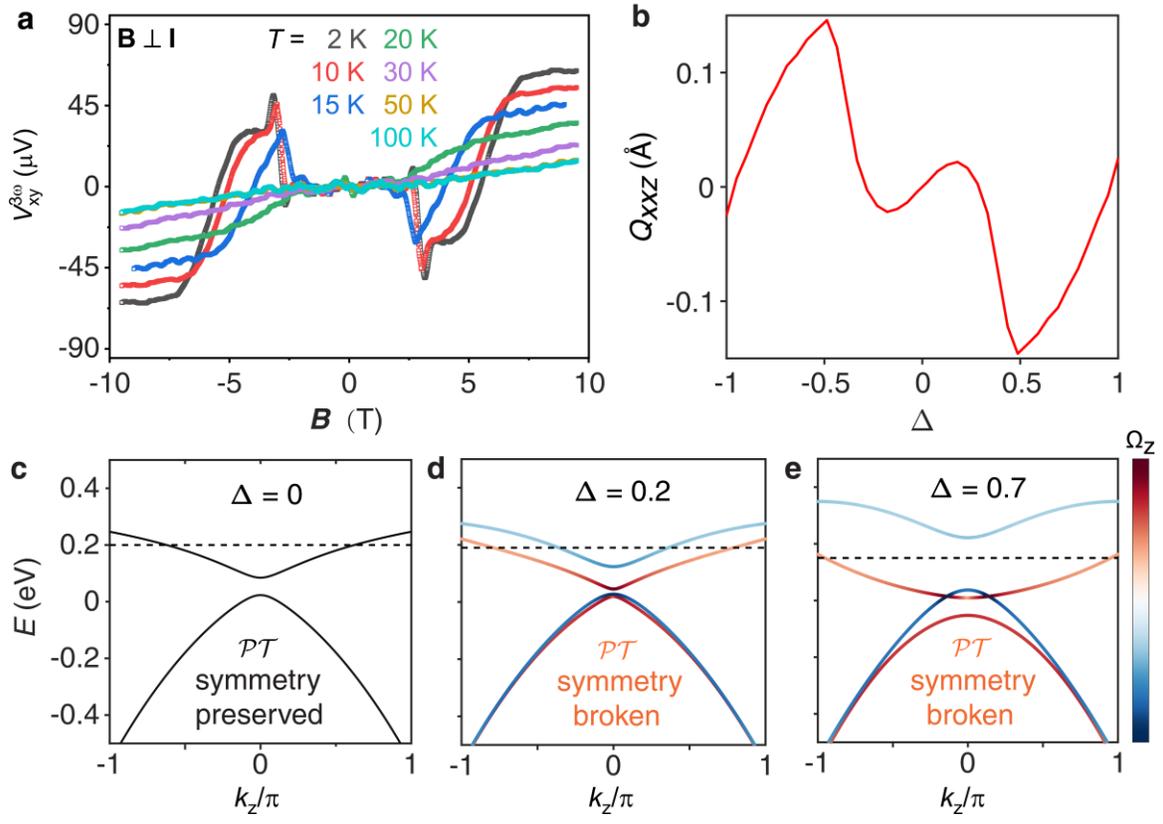

**Figure 4**